\documentclass{article}
\usepackage[utf8]{inputenc}
\usepackage{authblk}
\usepackage{indentfirst}
\usepackage{graphicx}
\usepackage{float}
\usepackage{amsmath, amssymb, txfonts}
\usepackage{appendix}

\usepackage[width=.75\textwidth]{caption}
\newtheorem{theorem}{Theorem}[section]
\newtheorem{corollary}{Corollary}[theorem]

\newtheorem{definition}{Definition}[theorem]

\title{\textbf{General Compound Hawkes Processes\\ for Mid-Price Prediction}}

\author[1]{Myles Sjogren}
\author[2]{Timothy DeLise}

\affil[1]{Department of Mathematics and Statistics, University of Calgary \protect\\ 2500 University Drive NW, Calgary, AB, Canada T2N 1N4
\protect\\  myles.sjogren@ucalgary.ca \\ 
} 

\affil[2]{Department of Mathematics and Statistics, 
Université de Montréal  \protect\\
Pavillon André-Aisenstadt (AA-5190)
2920, chemin de la Tour \protect\\
timothy.delise@umontreal.ca
}

\date{October 2021}
\begin{document}

\maketitle

\abstract{
High frequency financial data is burdened by a level of randomness that is unavoidable and obfuscates the task of modelling. This idea is reflected in the intraday evolution of limit orders book data for many financial assets and suggests several justifications for the use of stochastic models. For instance, the arbitrary distribution of inter arrival times and the subsequent dependence structure between consecutive book events. This has lead to the development of many stochastic models for the dynamics of limit order books. In this paper we look to examine the adaptability of one family of such models, the General Compound Hawkes Process (GCHP) models, to new data and new tasks. We further focus on the prediction problem for the mid-price within a limit order book and the practical applications of these stochastic models, which is the main contribution of this paper. To this end we examine the use of the GCHP for predicting the direction and volatility of futures and stock data and discuss possible extensions of the model to help improve its predictive capabilities.} \\

$\mathbf{Keywords}$: Limit order book,  Hawkes process, Futures data, Diffusive limit, Price prediction

\section{Introduction}

World financial exchanges are highly integrated and accessible to almost anyone in the world. The electronic limit order book has become a fundamental way to understand, interact with and engineer financial markets. Limit order books are in fact the main driver and source of data for high-frequency traders\cite{Ntakaris_2018}. As such it has become a point of interest for research in machine learning\cite{Zhang_2019} and stochastic modeling\cite{risks8010028}.

Limit order books are characterized by an expanding range of price \textit{levels} and \textit{volumes}. The collection of all quoted prices for a particular asset form a discrete grid where the space separating the possible prices is known as the tick size. At any given time one can place an order to buy or sell an asset at a desired price and volume in a variety of ways. The most common way is through limit orders which are called \textit{bids} if they are buy orders below the current market price or \textit{asks} if they are sell orders above the current market price. For example, a level 1 ask (best ask) is the lowest price amoung all outstanding sell orders. The level 2 ask is then the next highest price with sell orders where the minimum distance between the level 1 and level 2 ask is 1 tick.

The main divergence here from traditional stochastic modeling of assets prices is that, in limit order books, the price is viewed as a discrete (not continuous) process. We don't assume that the price gradually changes, but rather jumps from one price to another on the grid separated by multiples of the defined tick size. In fact, in the limit order book, there is no single price of an asset, but instead there are only orders. For this purpose the \textit{mid-price} is often modeled as an indication of the current price\cite{Ntakaris_2018}. This too is a discrete process since the mid-price is just the average of the best bid and best ask. In this way there is a fundamental difference to understanding the price process via limit order books as opposed to more traditional methods, such as Itô processes\cite{steele2012stochastic, bjork}.

Existing models\cite{steele2012stochastic} of price processes as smooth continuous processes begin to disconnect with reality at the micro perspective. Exchanges that actively use limit order books process orders at discrete price levels, and thus there is no real "price", but instead of history of past transactions and outstanding, unfilled orders. Transactions take place at discrete price levels, so the time series evolution of the transactions exhibits jumps, the effect of which become more profound at smaller time scales. In the past some academics have advocated for the use of continuous time models for market micro-structure, and subsequently introduced a framework based on point processes \cite{Engle1998s,Hasbrouck1991}. Since this time, the study of point processes, and in-particular Hawkes processes, have been the focus of an increasing amount of academic study\cite{bacry2015hawkes,laub2015hawkes}.

A recent application of Hawkes processes in finance is the development of a new type of stochastic model named the General Compound Hawkes Process (GCHP)\cite{swishchuk2017general,Swishchuk2021}. This model was first introduced to model the risk process in insurance, but has been adapted for other problems such as modelling the mid-price process of a limit order book. This model consists of two main components; firstly a Hawkes process is used to model the number and timing of new events over a given time interval; secondly a Markov chain is used to predict the type of the next event where the type of the next event is associated with a given state of the Markov chain and pertains to the size and direction of the predicted movement. Simple models restrict potential movements to be fixed up and down values while more complex models define different states for the Markov chain using a quantile based approach on the collection of all observed price movements. Using asymptotic methods this model can demonstrate a link between order-flow and volatility which can be used to asses the fit of the model. Further the volatility of the price changes can be expressed in terms of parameters describing the rate of arrivals and price changes.

In this paper we look to study the use of this General Compound Hawkes Process for new types of financial data and analyze its predictive capabilities. To this end we apply the model to futures data as a means to affirm existing theory for this new type of financial data. In doing this we can confirm certain universal empirical peculiarities of financial data such as the clustering of events arrivals and the existence of different volatility regimes over different time frames. Previous literature has stopped short of implementing prediction procedures for prices so also we seek to apply existing theory to predict mid-price movements. We implemented two distinct methods of prediction, the first uses diffusive limit theorems of the mid-price process and the second uses a Monte Carlo-esqe approach that entails aggregating many different simulated mid-price paths. In any case, at any given point in time we must only use past data to determine the models parameters and then use this calibrated model to evaluate the following section of time. In such a way, we loosely fit our technique into the family of \textit{sequential investment strategies}. These are sequential algorithms that simulate real-life, where we can condition the probability of future events on a history of events leading up to the current time. This methodology is also reminiscent of the application of modern machine learning algorithms, where we wish to predict the future based on historical data.

While the GCHP model was able to predict volatility levels consistently to some reasonable level of accuracy, the directional predictions where found to be more middling. For all tests the overall prediction accuracy was always above a chance level prediction, but barely in some instances. The result of any given test was shown to be heavily dependent on the experimental set-up applied and the parameters used. Nevertheless the model showed strength in predicting the volatility in the mid-price process of several different assets and, with sufficient fine tuning, showed potential in predicting directional movements.

The rest of the paper is organized as follows. Data is described in section 2, the Hawkes process is introduced in section 3, and the different types of GCHPs are introduced in section 4. Justification for the use of the chosen models is presented in section 5. In section 6 we outline the procedures used for the prediction problem and present empirical results of various experiments. Section 7 provides conclusions as well as ideas for future work.

\section{Data Description and Exploration}
There are two types of data used in this paper, the first is a proprietary historical sugar (SB) futures data set and the second is public historical stock data set for a variety of stocks including AMZN, FB, MSFT and more as described in and retrieved from \cite{Cartea2015AlgorithmicAH}. These two asset classes are seemingly unrelated, but it can be shown that most financial assets share some universal features which work well with the proposed models involving Hawkes processes. The set of SB futures data consists of every individual update to the top ten levels of the limit order books price and volume levels over a several month period starting in January 2021. Every row in this data corresponds to an LOB update, that is every time one feature changes another row of data is generated. As an effect there can be many stagnant periods where prices won't update, but the volume available at a certain level of depth will. This is a side effect of the ability to cancel active orders or submit new orders to the limit order book on a free flowing basis. Similarly, the public stock data also contains ten levels of limit order book price and volume data over a one month window in November 2014 (18 to 20 days per stock). This data is updated after every market order hence, is less abundant than the futures data while still tracking the evolution of the mid-price.

\begin{figure}[H]
\centering
  \includegraphics[width=.9\linewidth]{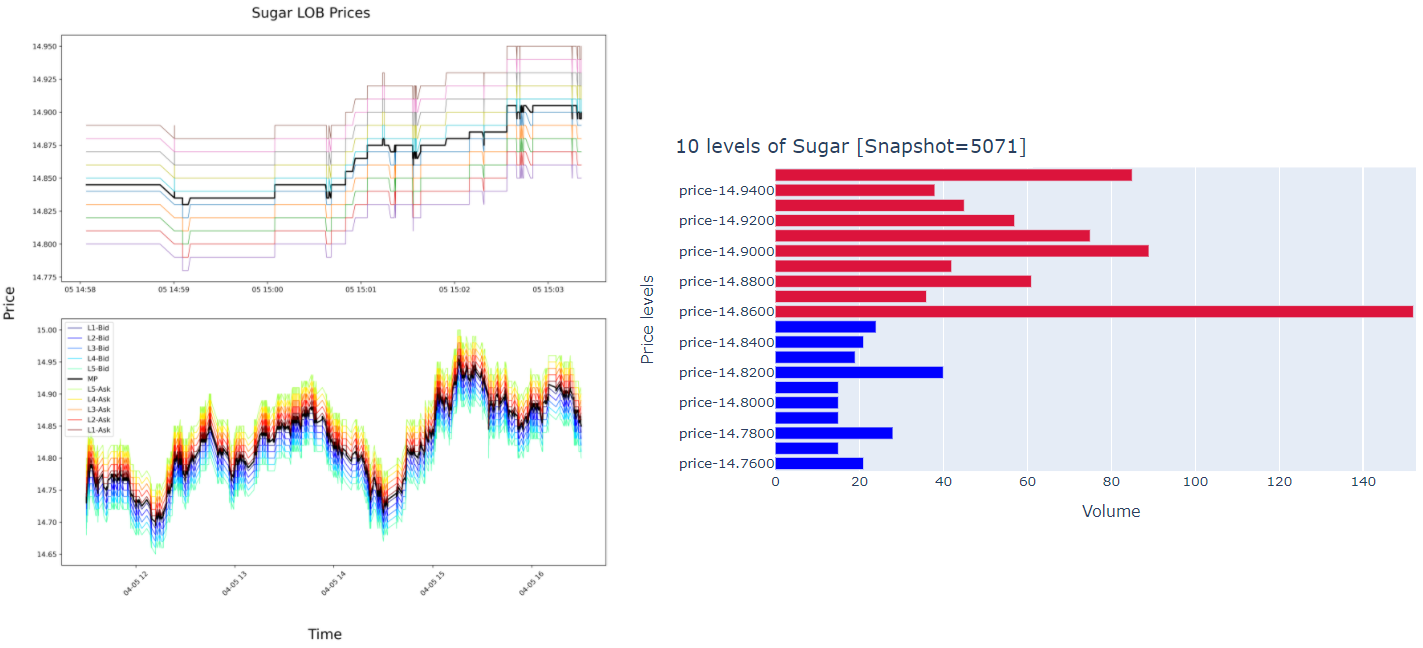}
  \caption{10 price levels (left) over a short time interval and 10 volume levels (right) of the sugar LOB taken at a select moment in time. Note that all the prices have a tendency to move simultaneously, but this isn't always the case.}
  \label{fig: SB LOB}
\end{figure}

 It is important to notice that the GCHP models, which will be explained more in detail later, are mid-price models. Each event in such a model corresponds to a mid-price change, either up or down. Each separate data-set also shows a different overall level of liquidity and different days can be more or less volatile than others. This fact can be demonstrated by looking at histograms for the average number of empirical mid-prices changes that occur over time intervals of fixed length for each separate data-set. The results of this for our sugar futures data are shown below in figure 2 and indicate that there is on average close to $2000$ mid-price changes per day and one about every dozen seconds. Comparing the same histograms for the AMZN stock data in figure 3 shows different results for the average number of mid-price changes with around $4000$ per day one one about every couple of seconds. Of note is that different underlying stocks have varying levels of liquidity as well as different distributions of the average number of mid-price changes. This suggests that the any model used in attempting in to model the mid-price must be calibrated to the individual data one is attempting to model. In other words, parameters for that fit one set of data will typically not fit another set of data for a different underlying asset. Another thing to note is that in spite of the frequency of updates to the limit order book in order to test the goodness of fit of the purposed model, only the first level of data is needed. Further the majority of price changes that occur are within one tick of the starting price. These things combined provide justification for a simpler model and helps combat the computational cost associated with processing the abundances of data. 
 
 \begin{figure}[H]
\centering
  \includegraphics[width=.8\linewidth]{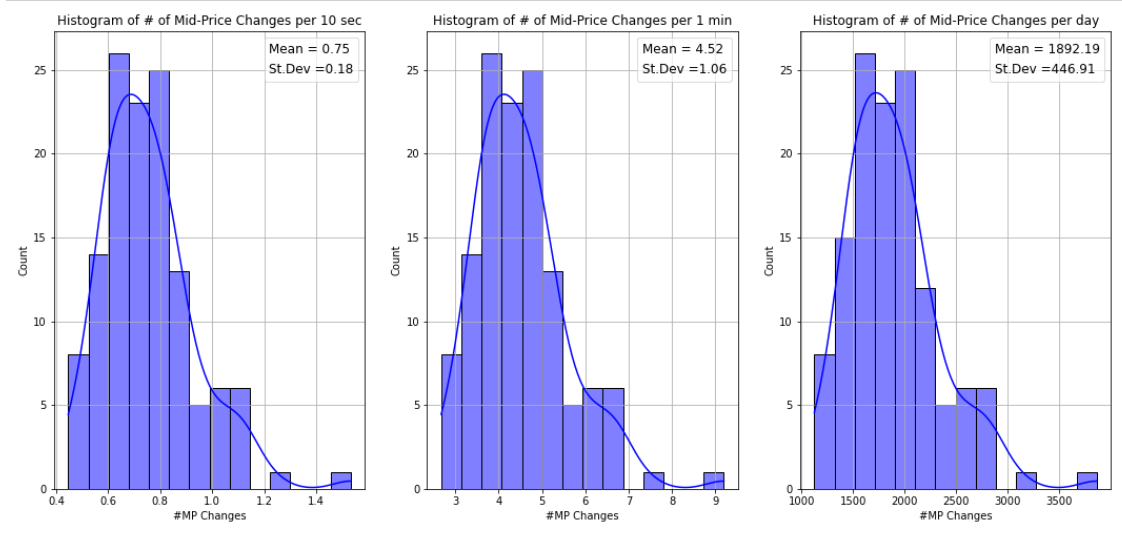}
  \caption{SB: Histograms for the average number of mid-price changes over specified time windows of 10 seconds, 1 minute and 1 day respectively from left to right.}
  \label{fig: SB_PC_hists}
\end{figure}

\begin{figure}[H]
  \centering
  \includegraphics[width=.8\linewidth]{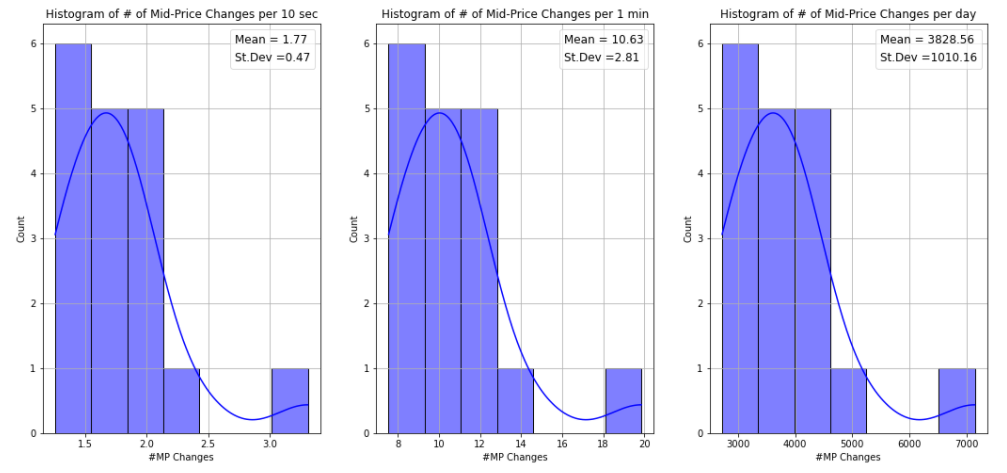}
  \caption{AMZN: Histograms for the average number of mid-price changes over specified time windows of 10 seconds, 1 minute and 1 day respectively from left to right.}
  \label{fig: AMZN_PC_hists}
\end{figure}
\noindent

\section{Hawkes Processes}
 
To build up to defining a Hawkes process we start by looking at counting processes which are time-dependent processes characterized by the arrival times of events, which in the context of this paper correspond to mid-price changes. \cite{daley2008,guo2020multivariate}.
 \begin{definition}
 	\textbf{Counting Process} A stochastic process is a counting process if it satisfies the following 3 properties.
	\begin{itemize}
  		\item $N(t) \ge 0,\ N(0) = 0$
  		\item $N(t+s) \ge N(t) , \ \forall t,s \ge 0$
  		\item $N(t)$ in an integer
	\end{itemize}
	
	$N(t+s)-N(t)$ is then the number of events occurring during the time interval $[t,t+s]$ and $N(t)$  is a right-continuous step function.
\end{definition}

A counting process can be viewed as a cumulative count of the number of event arrivals up to a certain time $t$, if $s < t$ then $N(t) - N(s)$ is the number of arrivals (counts) on the interval $(s,t]$. A well-known example of a counting process is the Poisson process, derived from the Poisson distribution\cite{kingma1993}. Examples of a counting processes would be the arrival of customers at a store or the arrival of orders to a financial market. Counting processes are one class of a more general class of processes called \textit{point processes}, which are subject of study concerned with the random distribution of points in space(time).

 \begin{definition}
 \textbf{Point Process} Let $(T_1,T_2,T_s,...)$ be a sequence of non-negative random variables with $P(0 \le T_1 \le T_2 \le T_3 \le ...) =1$, and let the number of points in a bounded region be finite almost surely, then $(T_1,T_2,T_s,...)$ is called a point process.
 The point process is characterized by the conditional \textbf{intensity function $\lambda(t)$} in the form 
$$\lambda(t) = \lim_{h \to 0} \frac{	E[N(t+h) - N(t)\text{   } |\text{   } \mathcal{F}^N (t)]}{	h}$$
where $\lambda(t)$ is a non-negative function and $\mathcal{F}^N (t)$, $t>0$ is the corresponding natural filtration. 
 \end{definition}

Point processes are a very general formulation. In nature we may find specific processes exhibiting  behaviour which at first glance may seemingly stem from a random distribution of arrivals, but after closer examination they may fail to be random at all. For example, if we want to understand the probability of an earthquake happening, this process exhibits non-random behavior. Instead such processes exhibit a \textit{self-exciting} behavior. One arrival increases the probability of the next. The following definition describes a \textit{Hawkes processes}, named after its creator to model seismic activity\cite{Hawkes1971}.\\

\begin{definition}
\textbf{1-Dimensional Hawkes Process} We say that a counting processes $N(t)$ is a 1-Dimensional Hawkes process when the intensity function is in the following form.
$$\lambda(t) = \lambda - \int_0^t \mu(t-s)dN_s$$		
Here $\lambda$ is a positive constant called the \textbf{background intensity} and $\mu(t)$ is the \textbf{excitation function} from $\mathbb{R}^+ \to \mathbb{R}^+$ with $\int_0^{+\infty} \mu(s)ds < 1$.
\end{definition}\

Overwhelmingly existing literature has been concerned with the exponential excitation function, also called the exponential kernel, due to its analytical tractability. This model sets $\mu(t) = \alpha e^{-\beta t}$, and so the Hawkes processes \textit{with exponential decay} is thus a 3-parameter model which approximates a self-exciting point processes.

\begin{equation}\label{HawkesExponential}
	\lambda(t) = \lambda - \int_0^t \alpha e^{-\beta (t-s)}dN_s
\end{equation}

The usual way to fit the parameters of such a processes is by using algorithms such as particle swarm optimization to maximize the likelihood of a given set of data for the model.

\begin{theorem}\textbf{Hawkes Process Likelihood}\cite{laub2015hawkes,daley2008}\label{hawkeslike}
Let $N(\cdot)$ be a regular point processes on $[0,T]$ for some finite positive $T$ and let $t_1, t_2, ..., t_k$ denote a realization of $N(\cdot)$ on $[0,T]$. Then the likelihood $L$ of $N(\cdot)$ give a Hawkes process with intensity function $\lambda(t)$ is expressed in the following form.
$$L=\left[ \prod_{i=1}^k \lambda(t_i) \right]\exp\left( - \int_0^T \lambda(u) du \right)$$
\end{theorem}

Here we should note that theorem \ref{hawkeslike} does not depend on a particular excitation function $\mu(t)$ and this detail motivates it's use for fitting more general Hawkes processes than those with exponential decay. In practice the log-likelihood is often preferred which was introduced in \cite{Ogata1978} as follows.
\begin{equation}
	\log(L)=\sum_{i=1}^k \log\left(\lambda(t_i)\right) - \int_0^T \lambda(u) du
\end{equation}

\section{The GCHP Mid Price Models}
For the purpose of our experiments we have implemented four variations of a mid-price model based on the Hawkes process. These models differ only in the number of states used to characterize the size and direction of potential price changes; the basis of the following definitions is explored in \cite{swishchuk2017general, SwishchukVadori}.

\subsection{Models}

\begin{definition}
\textbf{General Compound Hawkes Process with Dependent Orders(GCHPDO)} Let the mid-price of the underlying be given by,
$$S_t = S_0 + \sum\limits_{k=1}^{N(t)} X_k$$
where $N(t)$ is a Hawkes process and $X_k \in \lbrace +\delta, -\delta \rbrace$ is a Markov chain where $\delta$ is fixed to the tick size of the underlying asset. Then we say the process defined by $S_t$ is a General Compound Hawkes Process with Dependent Orders.
\end{definition}\

\begin{definition}
\textbf{General Compound Hawkes Process with 2 State Dependent Orders(GCHP2SDO)} Let the mid-price of the underlying be given by,
$$S_t = S_0 + \sum\limits_{k=1}^{N(t)} a(X_k)$$
where $N(t)$ is a Hawkes process, $X_k$ is an ergodic continuous time Markov chain independent of $N(t)$ with state space $ X = \lbrace 1,2 \rbrace$ and $a(.)$ is a continuous and bounded mapping function on X such that, 
\begin{itemize}
    \item $a(1)$ is the mean off all upward price movements
    \item $a(2)$ is the mean of all downward price movements
\end{itemize}  Then we say the process defined by $S_t$ is a General Compound Hawkes Process with 2 State Dependent Orders.
\end{definition}\ 

\begin{definition}
\textbf{General Compound Hawkes Process with 4 Dependent Orders(GCHP4DO)} Let the mid-price of the underlying be given by,
$$S_t = S_0 + \sum\limits_{k=1}^{N(t)} a(X_k)$$
where $N(t)$ is a Hawkes process, $X_k$ is an ergodic continuous time Markov chain independent of $N(t)$ with state space $ X = \lbrace 1,2,3,4 \rbrace$ and $a(.)$ is a continuous and bounded mapping function on X such that,
\begin{itemize}
    \item $a(1)$ = mean of all upward price movements $\geq$ one tick
    \item $a(2)$ = mean of all upward price movements $\geq$ one half tick and $<$ one tick
    \item $a(3)$ = mean of all downward price movements $\leq$ one half tick and $>$ one tick
    \item $a(4)$ = mean of all downward price movements $\leq$ one tick
\end{itemize} 
then we say the process defined by $S_t$ is a General Compound Hawkes Process with 4 Dependent Orders.
\end{definition}\ 

\begin{definition}
\textbf{General Compound Hawkes Process with n State Dependent Orders(GCHPnSDO)} Let the mid-price of the underlying be given by,
$$S_t = S_0 + \sum\limits_{k=1}^{N(t)} a(X_k)$$
where $N(t)$ is a Hawkes process, $X_k$ is an ergodic continuous time Markov chain independent of $N(t)$ with state space $ X = \lbrace 1,2,3,..,n \rbrace$ and $a(.)$ is a continuous and bounded mapping function on X. Then we say the process defined by $S_t$ is a General Compound Hawkes Process with n State Dependent Orders.
\end{definition}\ 
Looking at these definitions together it should be noted that each of the first three defined models is a specific case of the GCHPnSDO with a select number of states and a specific state mapping function $a(.)$. There any many choices for the function $a(.)$, in the simpler defined models such as the GCHPDO it suffices to use $a(X_k) = X_k$. For higher state models, such as the GCHPnSDO, a quantile based approach is used to break up the collection of all price movements. The approach is outlined as follows:
\begin{itemize}
    \item Split the data into positive and negative price movements
    \item Compute evenly distributed quantiles on both sets of data and disregard duplicates
    \item Compute state values as the mean of all observations which fall between consecutive quantiles
    \begin{itemize}
        \item Assign state value $a(1)$ to any observation below the first quantile.
        \item Assign state value $a(i)$ to any observation above the i-th quantile and below the i+1-th quantile.
    \end{itemize}
\end{itemize}

 \begin{figure}[H]
  \centering
  \includegraphics[width=.5\linewidth]{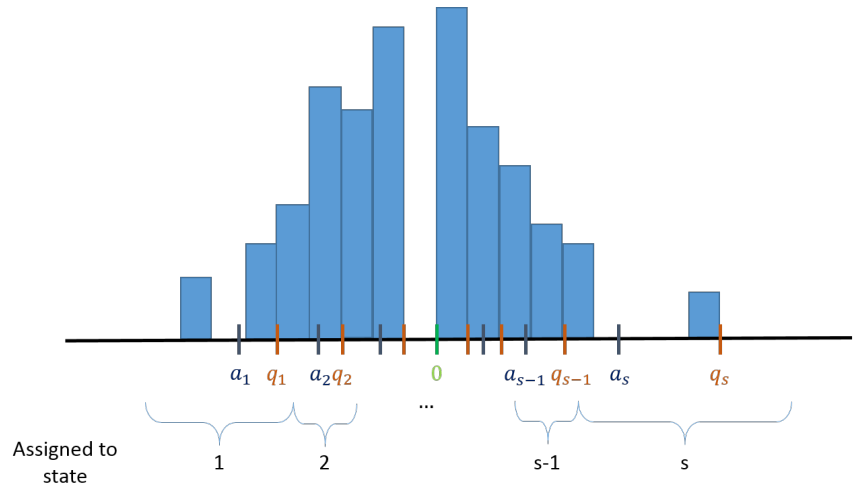}
  \caption{Illustration of Quantile Based Approach as seen in \cite{SwishchukHofmeister}.}
  \label{fig: quantile_method_pic}

\end{figure}

\indent These several variations on the general compound Hawkes process allow for increasing levels of flexibility in regards to allowable price changes. This gives an option to adjust for different levels of volatility in the underlying by altering the scope of possible movements. In particular given an asset known to have low volatility decreasing the number of states and the corresponding size of allowed price changes can give a better fitted model. Similarly increasing the size of possible movements for more volatile assets will yield a better fitted model overall.

\subsection{Limit Theorems}
Asymptotic analysis in the context of high frequency trading can pertain to time scales of just a few minutes or hours, as such we can look to apply this type of analysis on a daily time scale to approximate the true price process. In previous works asymptotic analysis was done on the GCHP that uncovered a link between price volatility and order-flow. This link is registered through the diffusive limit of the mid-price process. In particular it can be shown that this limit is expressible in terms of parameters of the Hawkes processes intensity function and parameters defined through the Markov chain that pertain to mid-price changes. For more details on the following theorems see \cite{risks8010028, SwishchukMerton}.
\begin{theorem}(Jump Diffusion Limit for GCHPnSSO) Let $X_k$ be an ergodic markov chain with n states $X = \lbrace 1,2,...,n \rbrace$ with ergodic probabilities $(\pi^*_1, \pi^*_2 , ...., \pi^*_n)$ and let $S_t$ be defined as in definition 4.0.4.  Then,
$$\frac{S(nt) - N(nt) a^*}{\sqrt{n}} \xrightarrow{\text{n} \to \infty} \sigma^* \sqrt{\frac{\lambda}{1 - \hat{\mu}}} W(t)$$
where $W(t)$ is a standard Brownian Motion term and the others parameters are defined as follows:
\begin{itemize}
    \item $0 < \hat{\mu} := \int\limits_0^\infty \mu(s)ds \text{  }\text{ such that } \text{  }\int\limits_0^\infty s \mu(s)ds < \infty$
    \item $(\sigma)^2 := \sum\limits_{i \in X} \pi_i^* v(i)$
    \item $v(i) = b(i)^2 + \sum\limits_{j \in X} (g(j) - g(i))^2 p(i,j) - 2b(i) \sum\limits_{j\ in X} (g(j) - g(i))P(i,j)$
    \item $a^* := \sum\limits_{i \in X} \pi_i^* a(X_i)$
    \item $ b(i) := a(i) - a^* \text{  ,  } b = (b(1),b(2),...,b(n))'$
    \item $g := \left(P + \Pi^* - I \right)^{-1}b$
\end{itemize}
Here $P$ is the transition probability matrix for $X_k$ and $\Pi^*$ is a matrix containing the stationary probabilities of $P$. 
\end{theorem}

Note that here the transition probability $P(i,j) = P(X_{k+1} = j \text{  } | \text{  } X_k = i)$ corresponds to the $i,j$ entry of the matrix $P$ and that $\Pi$ can be approximated using $\Pi \approx \lim\limits_{n \to \infty} P^n$. This theorem suffices for each simpler version of the model but, more convenient expressions can be derived for the parameters in the instance that $n=2$, see \cite{risks8010028} for details. The proof of this theorem stems from variations on the LLN and CLT for both the Hawkes process and the GCHP. In general this theorem gives us a means to test the fit of a GCHP by comparing the volatility of empirical data to an associated theoretical diffusive limit. An alternative to this theorem that uses a pure diffusive limit independent of the number of states exists and can be used to aid in the long term prediction of the price process.

\begin{theorem} \label{thm:diff} (Pure Diffusion Limit for GCHPnSSO) Let $X_k$ be an ergodic Markov chain with n states $X = \lbrace 1,2,...,n \rbrace$ with ergodic probabilities $(\pi^*_1, \pi^*_2 , ...., \pi^*_n)$ and let $S_t$ be defined as in definition 4.0.4, $0 < \hat{\mu} := \int\limits_0^\infty \mu(s)ds$ and  $\int\limits_0^\infty s \mu(s)ds < \infty$. Then,
$$\frac{S(t) -  a^* \frac{\lambda}{1- \hat{\mu}} t}{\sqrt{t}} \xrightarrow{\text{t} \to \infty} \bar{\sigma} N(0,1)$$
where $N(0,1)$ is a standard normal CDF, $\bar{\sigma}$ is defined as follows:
$$\bar{\sigma} = \sqrt{(\sigma^*)^2 + \left( a^* \sqrt{\frac{\lambda}{(1 - \hat{\mu})^3}}\right)^2}$$
and $\sigma^* = \sigma \sqrt{\frac{\lambda}{(1- \hat{\mu})}}$ where $\sigma$, $a^*$ are defined as in the previous theorem.
\end{theorem}

\section{Model Justification} 
In the section we seek to justify the use of the GCHP for our data using methods first described in \cite{SwishchukVadori}. These methods show that the assumption of uniformity in event arrivals is a bad assumption because in reality event arrivals within a limit order book are often arbitrarily distributed instead. 
\subsection{Clustering Effect}
Due to the self-exciting intensity function of the Hawkes process it has a tendency to generate events in clusters. This is because when one event occurs it creates a burst in intensity which increases the subsequent likelihood of seeing another event immediately afterwords. The end result of this is clustered data and this pattern can be shown to exist in real data by plotting the average number of mid-price changes over one minute bars. This clustering pattern is shown in figure 5 for four consecutive days of sugar data.
\begin{figure}
	\centering
    \includegraphics[width=0.75\textwidth]{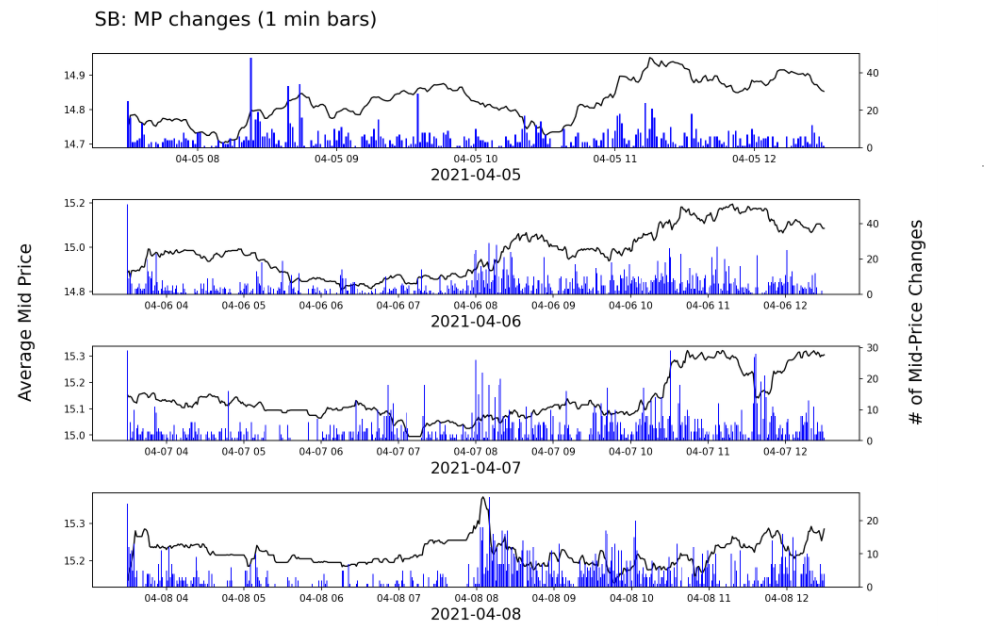}
    \caption{In this figure we see evidence of a clustering effect in the number of price changes on four consecutive days of sugar futures data. We also see evidence that large price swings are associated with more densely compacted clusters of price changes.}
    \label{Fig: Clustering}
\end{figure}

\subsection{Arbitrary Distribution of Inter-Arrival Times}
In the past a prevailing assumption was that events in a limit order book arrive according to a Poisson process and as such their inter-arrival times were exponentially distributed. This assumption can be used to define simple models with high levels of tractability, but in more recent times it has been shown to be a bad assumption given that inter-arrival times follow arbitrary distributions. This in turn advocates the use of a Hawkes process as with a Hawkes process the rate of arrivals is heavily influenced by the timing and number of recent events which directly leads to non-standard distributions. To verify this concept for our data we can plot the empirical cdf of inter-arrival times for the sugar data against theoretical cdf's for various distributions such as Exponential, Gamma, Weibull, Reciprocal and Wald distributions which are known to have similar shapes to the empirical cdf. In figure 6 we see that on different days the best fitting distribution can often be Weibull or Gamma, but it was found to never be the case that the best fitting distribution was exponential. 
 \begin{figure}
  \centering
  \includegraphics[width=.6\linewidth]{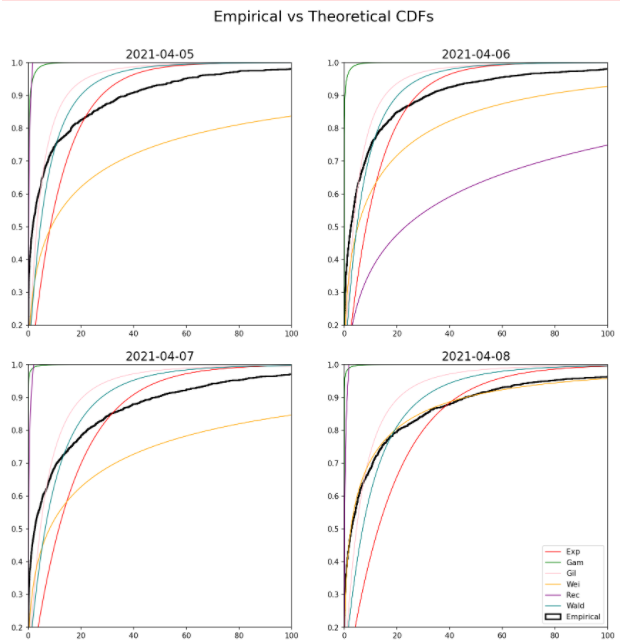}
  \caption{Empirical vs various Theoretical cumulative distribution functions for four days of sugar data.}
  \label{fig: CDf's}
\end{figure}

\subsection{Dependence of Consecutive Book Events}
Another aspect of this topic to consider is how dependent one book event is on the previous book event and to a greater extent how long does the influence of one book event last. To explain this we can look towards the intensity function of the Hawkes process to see how large every spike in intensity created by an arrival is and at what rate that created intensity decays. However, to first establish the existence of a sophisticated dependence between consecutive events we can compute the auto-correlation of the number of price changes over set time intervals. In particular we can use the following expression to quantify the dependence of mid-price changes,
$$C(\tau,\delta) = \frac{\mathbb{E}\left[(N_{t+ \tau} - N_t)(N_{t + 2\tau + \delta} - N_{t+ \tau + \delta})\right] - \mathbb{E}\left[(N_{t+ \tau} - N_t) \right] \mathbb{E}\left[(N_{t + 2\tau + \delta} - N_{t+ \tau + \delta})\right]}{\sqrt{ var(N_{t+ \tau} - N_t) var(N_{t + 2\tau + \delta} - N_{t+ \tau + \delta})}}$$
where $\tau$ is the length of the time interval considered and $\delta$ is the time lag between compared windows. This function compares the number of price change over an interval of length $\tau$ to increasingly lagged intervals of the same size. In the below figures we have computed this function for a single day of sugar data and compared different values for the parameter $\tau$.
 \begin{figure}
  \centering
  \includegraphics[width=.6\linewidth]{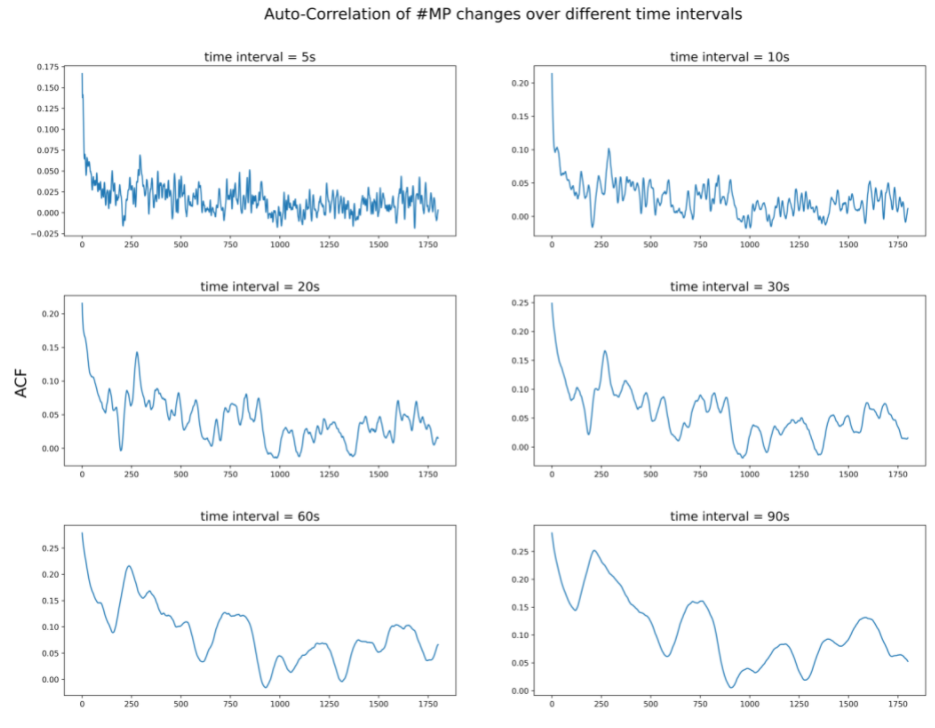}
  \caption{Auto-correlation function for selected time intervals for one day of sugar data.}
  \label{fig: acf}
\end{figure}
From figure 7 below we can see that for all values of $\tau$ tested there exists some level of positive correlation that decays over time and that after 15-30 minutes non-negligible positive correlations exist. So we can conclude that the impact of an event lasts for some time and that one event is not independent from the last. Both of these ideas together reiterate the potential use of a Hawkes process in modelling event arrivals.

\section{Prediction Process}

\subsection{Model Fitting and Selection Procedures} The best model for any given subset of data may vary based on the volatility intrinsic to said subset. For instance subsets of data taken for time windows closer to market closing tend to be more volatile overall. Consequently certain models may persistently under estimate the volatility of the data due to a higher frequency of price moves and a larger typical variation in the sizes of said moves. Choosing the best model for any given set of data is a process involving fitting each component of the defined model and can be broken down loosely into the following steps:
\begin{itemize}
    \item Fit a Hawkes Process to data and estimate the parameters of the arrival process.  
    \item Define transition probabilities between states via empirical frequencies of movements over training set. 
    \item Apply diffusive limit theorems to find theoretical volatility and compare to empirical volatility observed.
\end{itemize} The details of steps 1,2 and the origination of this comparative analysis can be found in \cite{risks8010028,He}. For the last step recall that in theorems $4.1, 4.2$ we defined diffusive limits for the mid-price process as modelled by the GCHPnSDO,
$$\frac{S_{nt} - N(nt) a^*}{\sqrt{n}} \xrightarrow{\text{n} \to \infty} \sigma^* \sqrt{\frac{\lambda}{1 - \hat{\mu}}} W(t)$$
A well fitted model should be able to approximate empirical volatility over larger set time intervals. Thus by multiplying each side of the above expression by $\sqrt{n}$, looking at intervals of size $[in, (i+1)n]$ and taking $t=1$ we can define ,
$$S^*_i =   S_{(i+1)n} - S_{in} - (N((i+1)n) - N(in))a^*$$
and it then follows that for large values of $n$,
$$std(S^*_i) \approx \sqrt{n} \sigma \sqrt{\frac{\lambda}{(1 - \hat{\mu})}}$$
Plotting the empirical standard deviations defined above for increasing time windows and comparing the theoretical limits gives us a way to judge the best fitting model. To quantify this we can preform a comparison to the hypothetical best fitting curve using a least squares regression. In particular squaring on both sides our expression for the empirical standard deviation give us, 
$$std(S^*_i)^2 \approx  n \sigma^2 \left( \frac{\lambda}{(1 - \hat{\mu})} \right)$$
where the right hand side  can be viewed as a linear regression function with respect to n which has coefficient $c = \sigma^2 \left( \frac{\lambda}{(1 - \hat{\mu})} \right)$. Thus we define an error rate by,
$$ \left| \frac{\sqrt{c} - \sigma^* \sqrt{\frac{\lambda}{1 - \hat{\mu}}} }{\sqrt{c}}\right| $$

\indent Implementing this process over active trading hours for a given day of sugar futures data gives varied results. The two examples shown below indicate that the best model can vary day-to-day and that certain models are more well-fitted in general than others. In particular the simpler models have a tendency to underestimate the empirical volatility while the more complex models can sometimes over estimate it. Fitting each defined GCHP model on one day of sugar futures data gives varied results, two examples of days in which the best fitting model is different are displayed in figure 8 and figure 9. Summary statistics for the fitting process over the entire collection of sugar data is then displayed in figure 10 below.\\
 \begin{figure}[H]
\centering
\begin{minipage}{.5\textwidth}
  \centering
  \includegraphics[width=.85\linewidth]{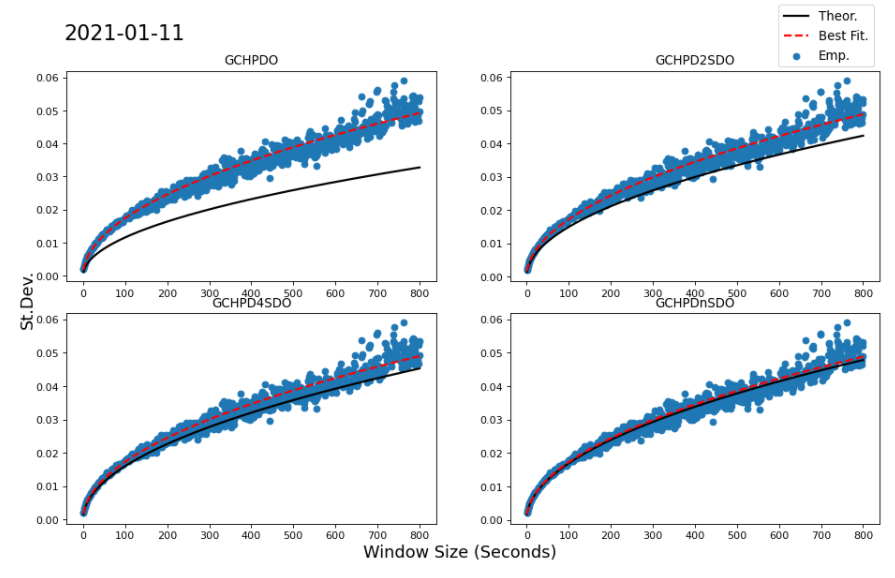}
  \caption{SB-Best fit : GCHPnSDO}
  \label{fig: SBFITD1}
\end{minipage}%
\begin{minipage}{.5\textwidth}
  \centering
  \includegraphics[width=.85\linewidth]{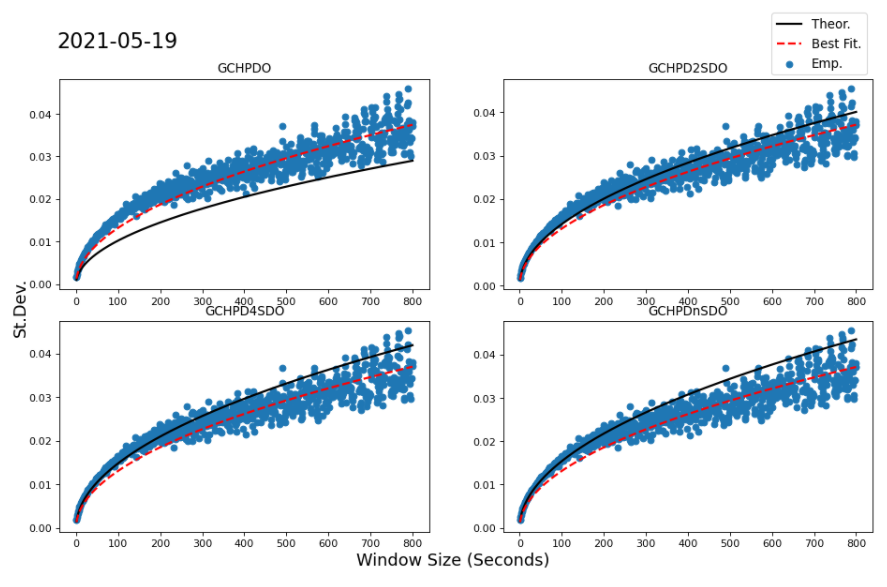}
  \caption{SB-Best fit : GCHP2SDO}
  \label{fig: SBFITD2}
\end{minipage}
\end{figure}\text{   }\\
 \begin{figure}[H]
\centering
\begin{tabular}{|c|c|c|c|c|c|}\hline
    & Best Fit's  & \% of Time Best Fit  & Mean Error Rate & Best Error Rate & Worst Error Rate  \\ \hline
    GCHPDO &  4 & 3.15 & 6.71 & 1.19 & 10.34\\
    GCHP2sDO &  33 & 25.98 & 5.44 & 0.02 & 16.26\\
    GCHP4DO & 26 & 20.47 & 2.77 & 0.05 & 13.05\\
    GCHPnSDO & 64 & 50.39 & 7.69 & 0.26 & 20.44\\ \hline
\end{tabular}
\caption{Summary statistics for model fitting process over all sugar data. In the table we see that the best model is consistently the n-state model as it is the best fitted model over half the time, however, it has a higher overall mean error rate across all days than the other models.}
  \label{fig: fit_stats_Sb}
\end{figure}\text{   } \\
For the collection of stock data we have less days to work with so the results are less likely to generalize, but the fitting process still produces some interesting results. For instance the model that fits best most often can change for distinct stocks and one model may never be chosen for a particular stock. For instance the AMZN stock data is not fit well by fixed tick models, but is well fit by adaptive tick models. Further analyzing the results of the fitting process it can be noted that days for which the best fitting model still gives a large error rate usually correspond to a day in which there was an abnormally large price change. Summary statistics for the fitting process on a few different stocks are presented in figures 13-15 below.

\begin{figure}[H]
\centering
\begin{minipage}{.5\textwidth}
  \centering
  \includegraphics[width=.85\linewidth]{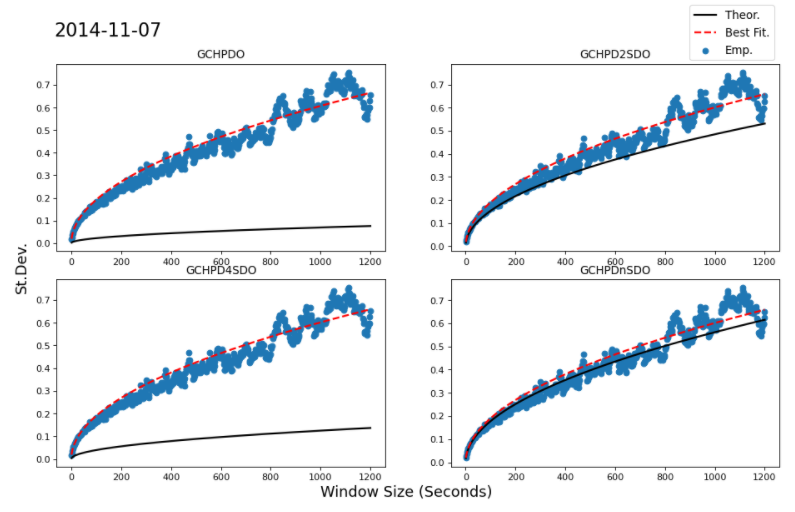}
  \caption{AMZN-Best fit : GCHPnSDO}
  \label{fig: AMZN curves}
\end{minipage}%
\begin{minipage}{.5\textwidth}
  \centering
  \includegraphics[width=.85\linewidth]{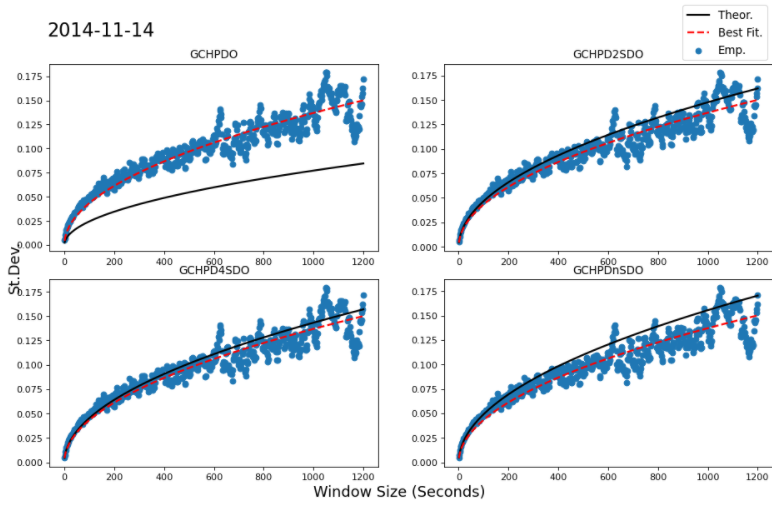}
  \caption{FB-Best fit : GCHP4DO}
  \label{fig: FB curves}
\end{minipage}
\end{figure}

 \begin{figure}
\centering
\begin{tabular}{|c|c|c|c|c|c|}\hline
    & Best Fit's  & \% of Time Best Fit  & Mean Error Rate & Best Error Rate & Worst Error Rate  \\ \hline
    GCHPDO &  0 & 0 & N/A & N/A & N/A\\
    GCHP2sDO &  3 & 16.7 & 4.16 & 1.92 & 6.85\\
    GCHP4DO & 0 & 0 & N/A & N/A & N/A\\
    GCHPnSDO & 15 & 83.3 & 16.21 &  1.78 & 9.88 \\ \hline
\end{tabular}
\caption{Summary statistics for model fitting process over all AMZN data.}
  \label{fig:fit_stats_AMZN}
\vspace{10.5pt}  
  \begin{tabular}{|c|c|c|c|c|c|}\hline
             & Best Fit's  & \% of Time Best Fit  & Mean Error Rate & Best Error Rate & Worst Error Rate  \\ \hline
    GCHPDO &  0 & 0 & N/A & N/A & N/A\\
    GCHP2sDO &  1 & 5.6 & 2.13 & 2.13 & 2.13\\
    GCHP4DO & 9 & 50.0 & 7.97 & 0.51 & 23.98\\
    GCHPnSDO & 8 & 44.4 & 10.66 & 0.37 & 25.05 \\ \hline
\end{tabular}
\caption{Summary statistics for model fitting process over all EBAY data.}
  \label{fig: fit_stats_EBAY}
  
\vspace{10.5pt}  
  \begin{tabular}{|c|c|c|c|c|c|}\hline
    & Best Fit's  & \% of Time Best Fit  & Mean Error Rate & Best Error Rate & Worst Error Rate  \\ \hline
    GCHPDO &  0 & 0 & N/A & N/A & N/A\\
    GCHP2sDO &  3 & 16.7 & 1.59 & 0.22 & 3.57\\
    GCHP4DO & 7 & 50.0 & 6.79 & 0.11 & 24.02\\
    GCHPnSDO & 8 & 38.9 & 6.80 & 2.17 & 17.26 \\ \hline
\end{tabular}
\caption{Summary statistics for model fitting process over all FB data.}
  \label{fig: fit_stats_FB}
\end{figure}

\subsection{Prediction using Diffusive Limit}
Now we seek to apply the GCHP to the practical problem of future price prediction with three main objectives.
\begin{itemize}
    \item accurately predict directional mid-price movements above a certain size.
    \item gauge potential volatility in mid-price process.
    \item minimize absolute error of final predicted price.
\end{itemize}
For the first objective we create a three class classification problem for mid-price movements by setting a fixed threshold $\alpha$ to describe upward and downward movements. That is any price movement larger that $\alpha$ will be classified as an upward movement, any price movement below $-\alpha$ will be classified as a downwards movement and any movements in between will be classified as stationary. This gives us another parameter to tweak as the optimal value of this threshold can vary based on various factors such as the length of the prediction window. For the second objective we create a 2-class classification problem similarly to the previous task, but we instead classify all price movements above a set threshold $\alpha$ in absolute value for the purpose of measuring potential volatility instead of direction. For the third objective we simply compare the predicted mid-price to the corresponding realized value using historical data and analyze the distribution of differences between true and predicted values.\\
\indent Another consideration is the amount of time used to train/calibrate the model parameters before testing. For our experiments we purpose to have each respective training and test set be contained within the same day as to avoid overnight price jumps. That is we don't use data from one day to predict price movements several weeks later as it is known that the intensity created by book events is decaying in nature. Further we use a kind of walk forward validation scheme where for each consecutive test, that is we pick up our training and testing window and move them forward one hour at a time until we hit the end of the day. 
 \begin{figure}[H]
  \centering
  \includegraphics[width=.7\linewidth]{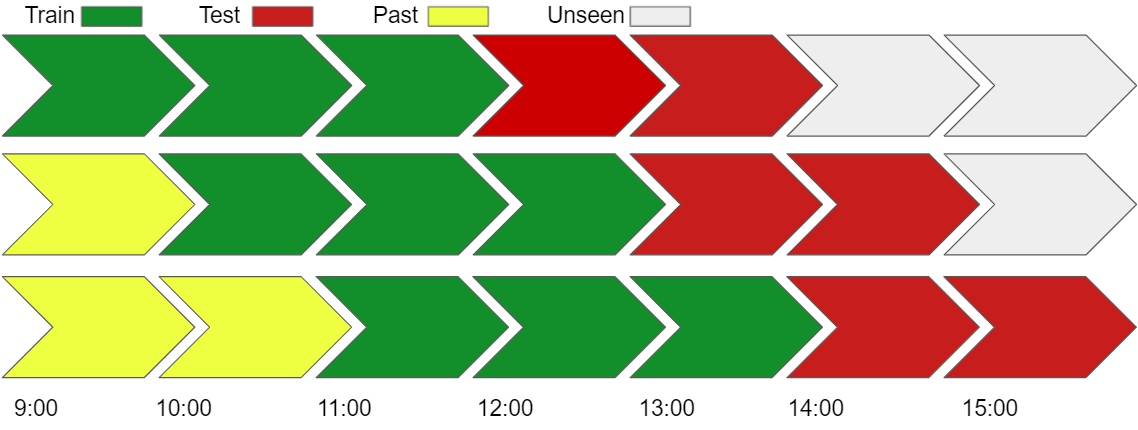}
  \caption{Illustration of 3 Hour/2 Hour train/test splits in walk forward scheme.}
  \label{fig: WFVS}
\end{figure}

\indent The first of the two methods we have used for prediction is based off the diffusive limit theorems presented in section four. That is we look to apply the defined diffusive limits of the mid-price process to make a prediction for the future mid-price at a set time past the end of the designated training window. So we calibrate the model parameters using historical data over a set time window and make a single numeric prediction for the future mid-price some fixed time after the end of said window. In particular we using to following corollary to theorem 4.2 as seen in\cite{SwishchukMerton}.

\begin{corollary} \label{cor:diff} Given the pure diffusive limit for the mid-price process stated in theorem 4.2 it follows that $S(t)$ can be approximated by,
$$ S(t) \approx S(0) + a^* \frac{\lambda}{1- \mathbb{\mu}} t + \bar{\sigma} W(t) $$
where $W(t)$ is a standard Brownian motion.
\end{corollary}
Based off this corollary, a more general prediction can be given by noting that the expectation of a standard Brownian motion is zero. Thus we propose to use the following approximation for the expected value of the mid-price on a large time interval.
$$\mathbb{E} [S(t)] \approx S(0) + a^* \frac{\lambda}{1- \mathbb{\mu}} t$$
\indent The following test was done for the sugar futures data using a 3 Hour, 2Hour train/test split. The thresholds for classes were set at $\alpha = 0.025$ for the 3-class problem and at $\alpha = 0.04$ for the two class problem. These values were chosen after extensive testing in order to create moderately balanced classes and give sufficient time for the limit theorem to take action.\\
\begin{figure}[H]
\centering
\begin{minipage}{.55\textwidth}
  \centering
  \includegraphics[width=1.05\linewidth]{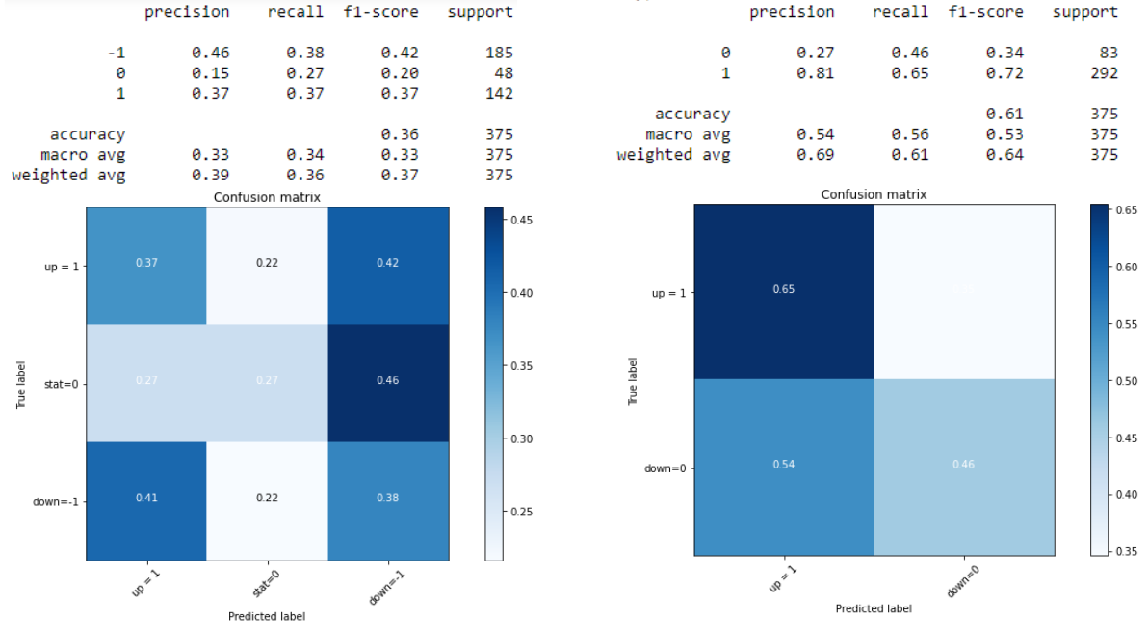}
  \label{fig: SB DL pred matrices}
\end{minipage}%
\begin{minipage}{.55\textwidth}
  \centering
  \includegraphics[width=.65\linewidth]{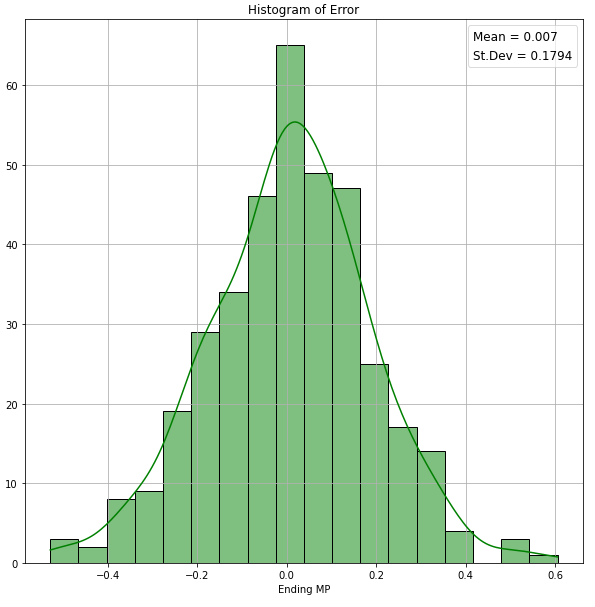}
  \label{fig: SB DL pred hist}
\end{minipage}
\caption{Summary results of 3-class classification problem and 2-class classification problem alongside histogram of total error for diffusive limit method on all sugar data. }
\end{figure}

From this test we can see that the overall predictive accuracy on the three class problem is only slightly above random chance, however, the accuracy for the volatility label is much better than random chance. This infers that this method of prediction shows a propensity for predicting the likelihood of future price movement, but not their direction. In both classification problems this methods preforms worse on the zero classes than the other classes which can be partially attributed to slightly imbalanced classes. We can also look at the error between the true realized value and the predicted value for the mid-price at the end of each testing interval. From the histogram in figure 18 we see that the differences between the true and predicted values of the mid-price across the whole data-set are normally distributed with mean approximately zero. So on average the final predicted value is close to the true value, but there are many cases where it can be off by many ticks as the standard deviation of these predictions is noticeably high.\\
\indent Now considering the stock data-set, given that each stock within it shows different levels of liquidity overall as well as different standards for average price changes, the hyper-parameters for the model should be specifically calibrated for each stock. This presents issues with the sample size of the stock data when running experiments over longer training/test intervals. Reported below are the results of a few experiments preformed for different stocks, the results are similar to that of the sugar futures data in that the model works best for the two class volatility label as compared to the three class label. The performance of this approach to prediction varied from stock to stock, for instance the results on the AMZN stock data were promising as the overall prediction accuracy on the two class label peaked over sixty percent.\\

\begin{figure}[H]
\centering
\begin{minipage}{.55\textwidth}
  \centering
  \includegraphics[width=1.2\linewidth]{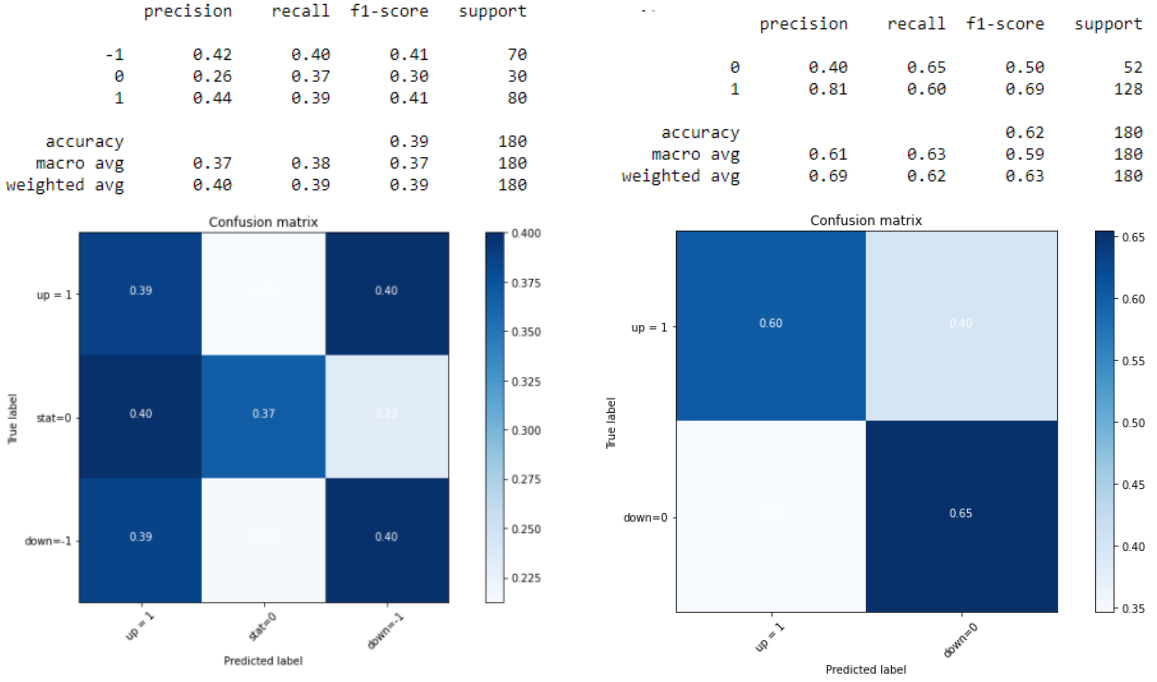}
  \label{fig: AMZN pred matrices}
\end{minipage}%
\begin{minipage}{.55\textwidth}
  \centering
  \includegraphics[width=.75\linewidth]{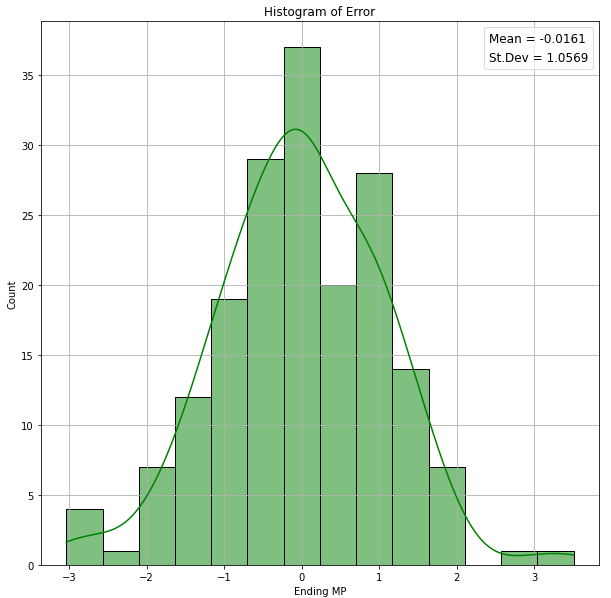}
  \label{fig: AMZN pred hist}
\end{minipage}
\caption{Predictions using diffusive limit technique for AMZN stock using a one hour/half hour train test split and thresholds $\alpha = 0.15$ and $\alpha = 0.3$ for the 3-class and 2-class label respectively. From the matrices on the left we can see that the again model struggled to predict the stationary classes in both cases, but the model showed a higher level of accuracy in its predictions on the other classes. The histogram of error for predictions is slightly positively skewed indicating that in this test the model had a tendency to over-estimate the true value of the mid-price.}
\end{figure}
\begin{figure}[H]
\centering
\begin{minipage}{.55\textwidth}
  \centering
  \includegraphics[width=1.0\linewidth]{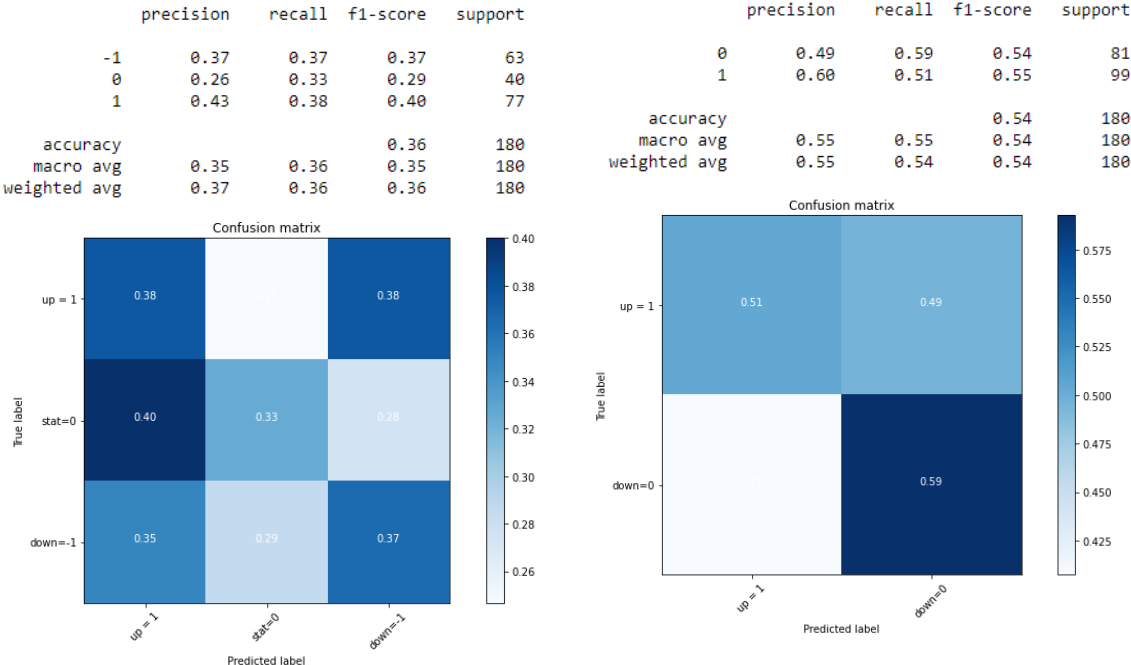}
  \label{fig: FB pred matrices}
\end{minipage}%
\begin{minipage}{.55\textwidth}
  \centering
  \includegraphics[width=.6\linewidth]{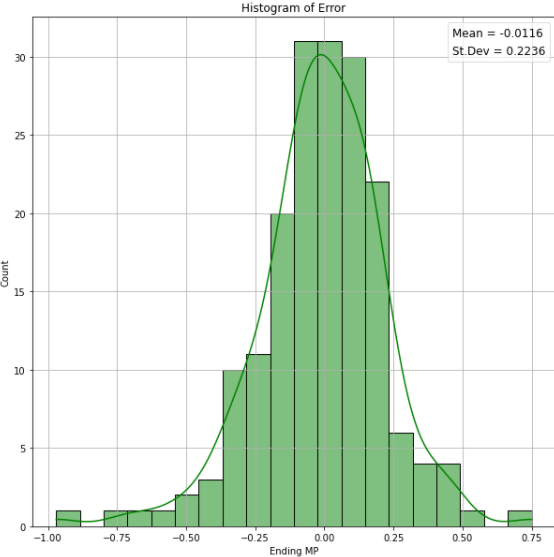}
  \label{fig: FB pred hist}
\end{minipage}
\caption{Predictions using diffusive limit technique for FB stock using a one hour/half hour train test split and thresholds $\alpha = 0.04$ and $\alpha = 0.08$ for the 3-class and 2-class label respectively.}
\end{figure}

\subsection{Prediction using Repeated Simulation}
An alternative approach to conducting prediction tasks is to simulate many paths of the underlying mid-price and use an aggregation of these many paths to form a final prediction. To accomplish this there are a few steps, first we simulate a Hawkes process over the testing interval to yield the timing of each consecutive event. Then we simulate the direction/size of each of these generated price changes based on the state space of the best fitting GCHP model over the current training interval. That is we maintain the current state of the model based on the previous state corresponding to the previous observed mid-price change. Conditioned on the current state we then simulate the discrete distribution of transition probabilities taken from the corresponding row of the empirical transition matrix. To form the final predicted mid-price we take the mean of the end points of all simulated paths. Below is an example of a collection of simulated mid-price paths for one test on the sugar data. We can observe that no one path exactly follows the realized path, but through repeated simulation we can attempt to predict the direction of the mid-price by capturing the general trajectory of all the simulated paths.
 \begin{figure}[H]
    \begin{center}
    \includegraphics[width=.38\textwidth]{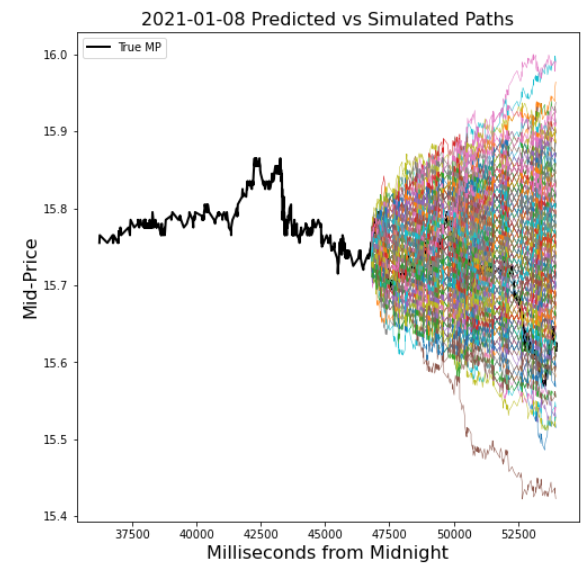}
    \end{center}
    \caption{Example of simulated paths for one test  window using sugar data.}
\end{figure}
Using this method introduces a new parameter in the number of simulated paths generated, this affects both the computational cost of running tests and has the potential to introduce mean reversion into the results. To elaborate, this method is burdened by an inherent component of randomness that can be mitigated by increasing the number of simulated paths used in creating the final prediction. As such running the same test multiple times will produce slightly different results every time, but it was found that results would only tend to vary on average by a couple of percentage points. There are both pros and cons to increasing and decreasing the number of simulated paths generated. Overall after various tests it was found that this method can often struggle to capture large price swings and trend reversals. As such the train/test split can drastically affect the results, but to gather a means of comparison to the diffusive limit method we keep all other parameters the same and report the results of the sugar test in figure 22 below.\\
\begin{figure}[H]
\centering
\begin{minipage}{.5\textwidth}
  \centering
  \includegraphics[width=1.1\linewidth]{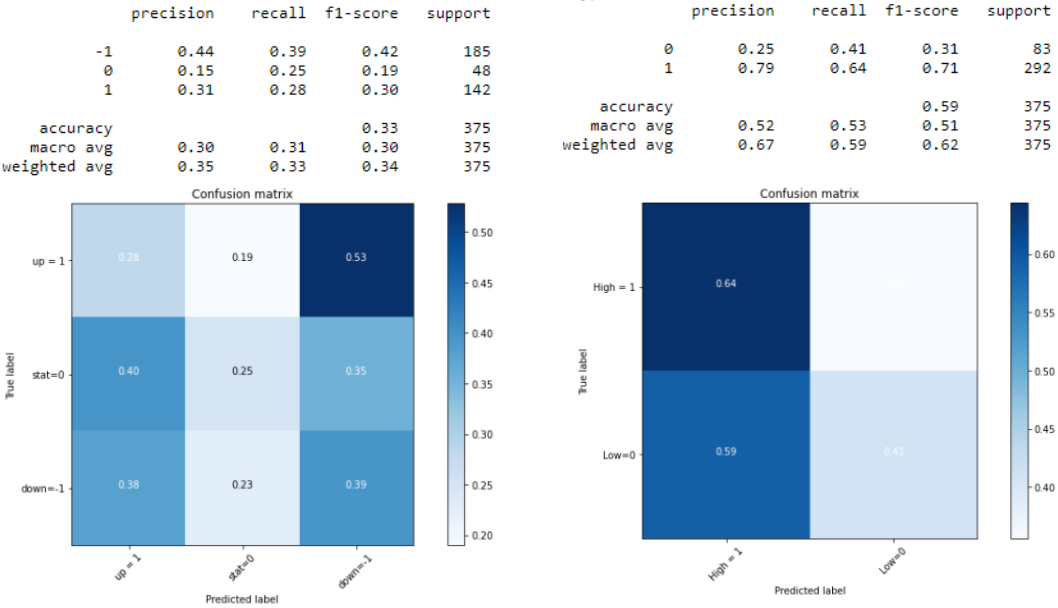}
  \label{fig: SB MC pred matrices}
\end{minipage}%
\begin{minipage}{.5\textwidth}
  \centering
  \includegraphics[width=.7\linewidth]{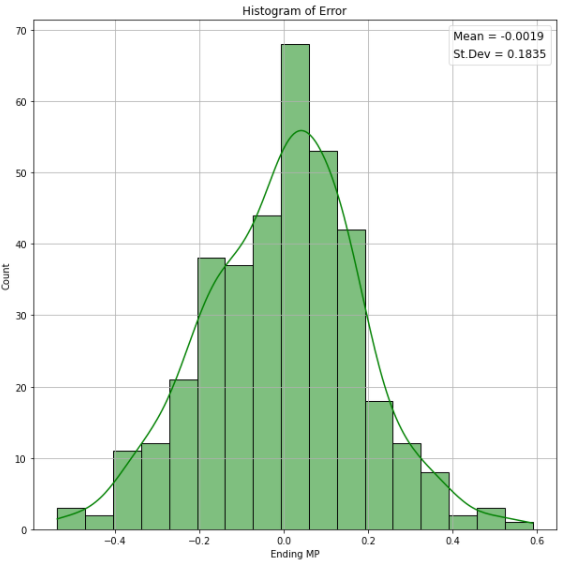}
  \label{fig: SB MC pred hist}
\end{minipage}
\caption{Summary results of using repeated simulation, a 3hour/2hour train/test split and the same class thresholds as with the diffusive limit test for all sugar data. This test used 250 simulated paths to determine the final prediction.}
\end{figure}

\
\section{Conclusions and Future Work}
Form this work we have established that the Hawkes processes embodies many empirical facts about financial data by reproducing justifying figures for a new type financial data. As such their use in the modelling of mid-prices processes is well warranted, but the Markov chain component of GCHP was shown to be less so. We were able to affirm existing theory on the GCHP model by applying the model fitting process to a more extensive amount of data. From this we have shown that the best fitting model is highly varied according to the underlying asset, time window fitted and more. For less liquid subsets of data the average size of mid-price changes tends to be smaller and most movements are subject to reversion. In this case a smaller number of states is required to best fit the model. On the other hand more liquid subsets of data with more inherent volatility require a larger number of states to best fit the GCHP. \\
\indent Despite being able to capture the dynamics behind the frequency, number and volatility of mid-price changes through the outlined model fitting process, when used for prediction the GCHP ran into various hindrances on performance. Firstly the computational cost of fitting and subsequently running experiments across substantial amounts of limit order book data was found to be quite large. In addition to this hyper-parameter tuning of the classification thresholds, length of training/testing data, number of simulated paths and more was found to be essential in producing decent results. Consequently finding the best possible experimental set-up is a time consuming endeavour which directly affected the results of all reported prediction tasks. In spite of this certain tests showed encouraging results in the directional predictions with overall accuracy peaking at around forty percent. However, the GCHP model preformed better for the volatility prediction suggesting that the model is much better at predicting the  likelihood of large price movements as compared to their direction. \\
\indent Future work on this topic firstly entails implementing a rigorous experimental set-up to determine the best hyper-parameters for the model across any given training/test window. This will directly contribute to improving the reported performance metrics and consequently the potential practical application of the model as a whole. It is also clear that using a single previous price movement as the states for a Markov chain in determining the next price movement doesn't fully capture the true dynamics of a series of book events. As such we have experimented with the use of higher order Markov chains in creating predictions though further work is needed in assessing the theoretical side of this approach as well as the computational challenges it introduces.  
\section{Acknowledgements}

This research was supported by Futures First and Mitacs through a collaborative research project.

\appendix
\section{Remark on the Expressivity of Predictions Using the GCHP Model.}

In corollary \ref{cor:diff} \cite{SwishchukMerton}, we describe an expression for the diffusive limit theorem \ref{thm:diff}.

$$ S(t) \approx S(0) + a^* \frac{\lambda}{1- \mathbb{\mu}} t + \bar{\sigma} W(t) $$

\noindent Here, a prediction about the direction of the future price movement is analogous to computing the expected value of $S(t) - S(0)$. If this is positive, we expect the future price to increase, and alternatively decrease if it is negative. 

$$\mathbb{E}[S(t)-S(0)] \approx \mathbb{E} \left[ a^* \frac{\lambda}{1- \mathbb{\mu}} t + \bar{\sigma} W(t) \right] $$

\noindent The Brownian motion term, $\bar{\sigma} W(t)$, is a martingale \cite{steele2012stochastic}, and as such its expected value is zero. The term $a^* \frac{\lambda}{1- \mathbb{\mu}} t$ is not random, and so its expected value is just the expression itself. These two steps allow the following simplification.

$$\mathbb{E}[S(t)-S(0)] \approx a^* \frac{\lambda}{1- \mathbb{\mu}} t  $$

\noindent In the right hand side of the above line, we notice the following, on the account that $\lambda$ and $t$ are positive and $0<\mu<1$ \cite{He}.
$$\frac{\lambda}{1- \mathbb{\mu}} t > 0$$

\noindent This is now to say that the direction of the future price is completely dependent on $a^*$. Now, recall that $a^*$ is defined as follows.
$$a^* := \sum \pi^*_i a(i)$$

\noindent In the case of the simplest 2 state model. We have $X=\{1,2\}$, $a(1)=1$, and $a(2)=-1$, yielding
$$a^* = \pi_1^* - \pi_2^* $$

\noindent and since $\pi_1^* + \pi_2^* = 1$, we finally have the result.

$$a^* = 2\pi_1^* - 1$$

\noindent We can now conclude that

$$\mathbb{E}[S(t)-S(0)] >0 \iff a^* > 0 \iff \pi_1^* > 0.5$$

\noindent The conclusion is that the direction of our prediction for any time in the future completely depends on the steady state probabilities of the Markov chain. The Hawkes process does not seem to be involved in the directional prediction. Let us recall that the steady state probabilities of a simple 2 state Markov chain are estimated as follows:

$$M = \begin{bmatrix}
1-p & p\\
q & 1-q
\end{bmatrix}$$
\noindent and consequently,
$$\pi^* = \left \{ \frac{p}{p+q}, \frac{q}{p+q} \right \}$$

\noindent so we finally we see that the steady state probabilities, and hence the entire directional prediction depend on the Markov chain transition probability matrix that is estimated from the data.

One take away is that if we base our steady state probability on a true estimate of the long-run probabilities, then our directional prediction will never change direction across all time. Hence, in this work, we have chosen to learn the Markov chain from a subset of recent history, but under this technique we have a momentum strategy, where our direction typically matches the direction of recent history. 

\section{Alternative Formulas for Predictions Using Diffusive Limits}

Based on the stated diffusive limit theorems 4.1, 4.2  \cite{He,risks8010028} there are some possible variations of note from the formula we used for mid-price predictions. Firstly using the pure diffusive limit, instead of taking the expected value we can instead include the Brownian motion term, this leads to the following alternative formulation,
\begin{align*}
    &S(t) \approx S(0) + a^* \frac{\lambda}{1 - \hat{\mu}}t +  \bar{\sigma} W(t)\\
    &\Longrightarrow s(t) \approx s(0) + a^* \frac{\lambda}{1- \hat{\mu}}t + \bar{\sigma} \sqrt{t} N(0,1) \\
\end{align*}
where  $W(t) \sim \sqrt{t} \text{ } N(0,1)$. We can also derive a formula from the jump diffusion limit which directly incorporates a simulated Hawkes processes into the prediction.
\begin{align*}
    &\frac{S(nt) - N(nt) a^*}{\sqrt{n}}  \to \sigma^* \sqrt{\frac{\lambda}{1 - \hat{\mu}}} W(t)\\
    &\Longrightarrow S(nt) - N(nt) a^*  \to \sigma^* \sqrt{nt} \sqrt{\frac{\lambda}{1- \hat{\mu}}}N(0,1) \\
    &\Longrightarrow  S(nt)  \to N(nt) a^*  + \sigma^* \sqrt{nt} \sqrt{ \frac{\lambda}{1- \hat{\mu}}}N(0,1) \\
    &\Longrightarrow  S(nt)  \approx S(0) +  N(nt) a^*  + \sigma^* \sqrt{nt} \sqrt{ \frac{\lambda}{1- \hat{\mu}}}N(0,1) \\
\end{align*}
In each of these methods we generate many predictions based off repeated simulations of a standard normal random variable and take the mean as the final prediction similar to section 6.3. Testing these methods gave slight variations in the results, as such they could be attributed to the new source of randomness. The most important change in performance metrics in each case there was a slight decrease in the mean as well as the standard deviation of the error in the final predictions.



\newpage
\bibliographystyle{abbrv}
\bibliography{main}
\end{document}